\documentclass{article}
\begin{document}
\title{Darboux's Theorem and Quantisation}
\author{Jos\'e M. Isidro \\ 
Department of Theoretical Physics,\\ 
1 Keble Road, Oxford OX1 3NP, UK\\ 
{\tt isidro@thphys.ox.ac.uk}}
\maketitle

\begin{abstract}

It has been established that endowing classical phase space
with a Riemannian metric is sufficient for describing quantum mechanics.
In this letter we argue that, while sufficient, the above condition is 
certainly not necessary in passing from classical to quantum mechanics. 
Instead, our approach to quantum mechanics is modelled on a statement that 
closely resembles Darboux's theorem for symplectic manifolds. 

Pacs codes: 03.65.-w, 03.65.Ca, 04.60.Ds.


Preprint no. OUTP-01-59P.

\end{abstract}

\tableofcontents

\section{Introduction}\label{intro}

Quantisation may be understood as a prescription to construct a quantum theory from a given 
classical theory. As such, it is far from being unique. Beyond canonical quantisation 
and Feynman's path--integral, a number of  different, often complementary approaches 
to quantisation are known, each one of them exploiting different aspects of the underlying 
classical theory. For example, geometric quantisation relies on the geometry of classical 
phase space. Berezin's quantisation can be applied to classical systems 
whose  phase space is a homogeneous K\"ahler manifold \cite{BEREZIN, SCHLICHENMAIER}. 

The deep link existing between classical and quantum mechanics has been known 
for long. Perhaps its simplest manifestation is that of coherent states \cite{COHST}. 
More recent is the notion that {\it not all quantum theories arise from quantising 
a classical system}. Furthermore, a given quantum model 
{\it may possess more than just one classical limit}.
These ideas find strong evidence in string duality and M--theory 
\cite{KAKU}.
 
It therefore seems natural to try an approach to quantum mechanics that is not based, 
at least primarily, on the the metric quantisation of a given classical dynamics. 
In such an approach one would not take a classical theory as a starting point. 
Rather, quantum mechanics itself would be more fundamental, in that its classical 
limit or limits (possibly more than one) would follow from a parent quantum theory.

\section{Berezin's metric quantisation}\label{berezin}

Below we briefly review the construction of the Hilbert space of states from the metric 
on complex homogeneous K\"ahler manifolds \cite{BEREZIN}.

Let $z^j$, $\bar z^k$, $j, k= 1,\ldots, n$, be  local coordinates on a
complex homogenous K\"ahler manifold ${\cal M}$, and let $K_{\cal M}(z^j,\bar z^k)$ be a K\"ahler
potential for the metric ${\rm d} s^2=g_{j\bar k}\,{\rm d} z^j{\rm d}\bar z^k$. The K\"ahler form
$\omega=g_{j\bar k}\,{\rm d} z^j\wedge {\rm d}\bar z^k$ gives rise to an integration measure
${\rm d}\mu(z,\bar z)$, 
\begin{equation} 
{\rm d}\mu(z,\bar z)=\omega^n={\rm det}\,(g_{j\bar k})\,\prod_{l=1}^n{{\rm d} z^l\wedge {\rm d}\bar
z^l\over 2\pi {\rm i}}.
\label{bermeasure}
\end{equation} 
The Hilbert space of states is the space ${\cal F}_{\hbar}({\cal M})$ of analytic
functions on
${\cal M}$ with finite norm, the scalar product being
\begin{equation}
\langle\psi_1|\psi_2\rangle =c(\hbar)\,\int_{\cal M}{\rm d}\mu(z,\bar z)\,{\rm
exp}(-\hbar^{-1}K_{\cal M}(z,\bar z))\,{\overline
\psi_1(z)}\psi_2(z),
\label{berscalar}
\end{equation}
and $c(\hbar)$ a normalisation factor.  Let  $G$ denote the Lie group of motions of ${\cal M}$, 
and assume $K_{\cal M}(z,\bar z)$ is invariant under $G$. Setting $\hbar=k^{-1}$, the family
of Hilbert spaces ${\cal F}_{\hbar}({\cal M})$ provides a discrete series of projectively unitary
representations of $G$. The homogeneity of ${\cal M}$ is used  to prove that the correspondence
principle is satisfied in the limit $k\to\infty$. Furthermore, let $G'\subset G$ be a maximal
isotropy subgroup of the vacuum state $|0\rangle $. Then coherent states $|\zeta\rangle $ are
parametrised by points $\zeta$ in the coset space $G/G'$.

\section{Quantum mechanics from Darboux's theorem}\label{statement}

Darboux's theorem locally trivialises any symplectic manifold:
every point of a $2n$--dimensional symplectic manifold 
possesses a local coordinate neighbourhood with coordinates $(p_l, q^l)$, $l=1,2,\ldots, n$,
in which the symplectic 2--form $\omega$ is expressed as $\omega={\rm 
d}p_l\wedge{\rm d}q^l$.

Let us now make the statement that

{\it Given any quantum system, there always exists a coordinate transformation 
that transforms the system into the semiclassical regime, i.e., into a system that 
can be studied by means of a perturbation series in powers of $\hbar$ 
around a certain local vacuum.}

As with Darboux's theorem and the Hamilton--Jacobi method \cite{ARNOLD},
one can see the use of coordinate transformations in order to trivialise 
a given system. In our context, however, trivialisation does not mean cancellation 
of the interaction term, as in the Hamilton--Jacobi technique. 
Rather, it refers to the choice of a vacuum around which to perform a perturbative 
expansion in powers of $\hbar$. As we will see presently, this is 
equivalent to eliminating the metric, thus rendering quantum mechanics 
metrically trivial. In this sense, Darboux's theorem for symplectic manifolds 
falls just short (by Planck's constant $\hbar$) of being a quantisation, 
as it otherwise provides the right starting point in the passage from classical 
to quantum mechanics. (In the strict sense of geometric quantisation 
\cite{WOODHOUSE}, only those symplectic manifolds that satisfy the 
integrality conditions can be quantised). Related geometric approaches to
quantum mechanics have been presented in refs. \cite{ASHTEKAR, ANANDAN, BFM, 
GQM}. More recently, ref. \cite{MEX} has brought to light the relevance 
for quantisation of infinite--dimensional, complex projective space.
Darboux's theorem has also been used in a very interesting approach to the 
quantisation of constrained dynamics in ref. \cite{JACKIW}; this latter approach 
is an effective alternative to Dirac's brackets.\footnote{I wish to thank 
R. Jackiw for drawing my attention to this point.} The classical 
equivalence between Dirac's method and that of Faddeev--Jackiw is 
established in ref. \cite{PONS}.

\section{Discussion}\label{discussion}

\subsection{The choice of a vacuum}\label{vuoto}

The statement above instructs us to choose a local vacuum.
Under the choice of a vacuum we understand a specific set of coordinates
around which to perform an expansion in powers of $\hbar$. This choice of 
a vacuum is local in nature, in that it is linked to a specific choice of 
coordinates. It breaks the group of allowed coordinate transformations to 
a (possibly discrete) subgroup, leaving behind a (possibly discrete) duality 
symmetry of the quantum theory. Call $q$ the local coordinate corresponding 
to the vacuum in question, and $Q$ its quantum operator. The corresponding 
local momentum $P$ satisfies the usual Heisenberg algebra with $Q$. 
This fact reflects, at the quantum level, the property that the 
Darboux coordinates $p$, $q$ render the symplectic form $\omega$ canonical,
$\omega={\rm d}p\wedge{\rm d}q$. However, as our starting point we have no 
classical phase space at all, and no Poisson brackets to quantise into commutators. 
This may be regarded as a manifestly non--perturbative formulation 
of quantum mechanics. 

\subsection{Quantum numbers {\it vs.} a topological quantum mechanics}\label{quantumnumbers}

Berezin's quantisation relied heavily on the metric properties of classical phase space. 
The semiclassical limit could be defined as the regime of large quantum numbers. 
The very existence of quantum numbers was a consequence of the metric structure.
If quantum mechanics is not to be formulated as a quantisation of a 
given classical mechanics, then we had better do away with global quantum 
numbers, {\it i.e.}, with the metric. Metric--free theories usually go by 
the name of topological theories. Hence  our quantum mechanics will be a
topological quantum mechanics, {\it i.e.} free of global quantum numbers.  
Locally, of course, quantum numbers do appear, but only after the choice 
of a local vacuum. 

\subsection{Classical {\it vs.} quantum} \label{versus}

After the choice of a local vacuum to expand around, the local quantum 
numbers one obtains describe excitations around the local vacuum chosen. 
Hence what appears to be a semiclassical excitation to a local observer
need not appear so to another observer. In fact may well turn out to be a 
highly quantum phenomenon, when described from the viewpoint of a different 
local vacuum. This point has been illustrated in ref. \cite{GQM},  
where coherent states that are local but cannot be extended globally 
have been analysed. The model of ref. \cite{GQM} provides an explicit example 
of the general procedure presented above.  

The logic could be summarised as follows: 1) the fact that this quantum mechanics 
is topological implies the  absence of
a metric; 2) the absence of a metric implies the absence of global quantum numbers;
3) the absence of global quantum numbers implies the impossibility of globally 
defining a semiclassical regime. The latter exists only locally.

\section{Summary}\label{suma}

In this paper we have analysed some general properties that quantum mechanics 
must satisfy, if it is not to be formulated as a metric quantisation of a given classical 
mechanics. We have formulated a statement, close in spirit to Darboux's theorem of 
symplectic geometry, that provides a starting point for a formulation 
of quantum mechanics that is explicitly metric--free. 
Our formalism may be understood as a certain limit of Berezin's 
quantisation. The latter relies on the metric properties 
of classical phase space ${\cal M}$, whenever ${\cal M}$ is a homogeneous K\"ahler manifold. 
In Berezin's method, quantum numbers arise naturally from the metric on ${\cal M}$. 
The semiclassical regime is then identified with the regime of large quantum numbers. 
Our method may be regarded as the topological limit of Berezin's quantisation, 
{\it topological} meaning that the metric dependence has been removed. 
As a consequence of this topological nature our quantum mechanics exhibits 
the added feature that quantum numbers are not originally present. 
They appear only after a vacuum has been chosen and, contrary to Berezin's
quantisation, they are local in nature, instead of global. 

It has been proved in ref. \cite{KLAUDER} that endowing classical phase space
with a Riemannian metric is sufficient for describing quantum mechanics.
In this letter we have argued that, while certainly sufficient, the above condition 
is not necessary in passing from classical to quantum mechanics.

{\bf Acknowledgments}

This work has been supported by a PPARC fellowship under 
grant no. PPA/G/O/2000/00469.

\end{document}